\documentclass[%
  aps,
  pre,
  superscriptaddress, letterpaper,
  longbibliography,
  showpacs,
  showkeys,
  amsfonts, amssymb, amsmath]{revtex4-1}

\usepackage{xcolor}
\usepackage[applemac]{inputenc}
\usepackage{epstopdf}							
\usepackage[caption=false]{subfig}
\usepackage{hyperref} 
\hypersetup{pdfpagemode={UseOutlines},
bookmarksopen=true,
bookmarksopenlevel=0,
hypertexnames=false,
colorlinks=true, 
citecolor=blue, 
linkcolor=red, 
urlcolor=black, 
pdfstartview={FitV},
unicode,
breaklinks=true,
}
\usepackage{graphicx,amssymb,amsmath}
\usepackage[english]{babel}

\begin{document}

\title{Transition from Weak Wave Turbulence to Soliton-Gas}
\author{Roumaissa Hassaini}
\author{Nicolas Mordant}
\email[]{nicolas.mordant@univ-grenoble-alpes.fr}
\affiliation{Laboratoire des Ecoulements G\'eophysiques et Industriels, Universit\'e Grenoble Alpes, CNRS, Grenoble-INP,  F-38000 Grenoble, France}

\begin{abstract}
We report an experimental investigation of the effect of finite depth on the statistical properties of wave turbulence at the surface of water in the gravity-capillary range.  We tune the wave dispersion and the level of nonlinearity by modifying the depth of water and the forcing respectively. We use space-time resolved profilometry to reconstruct the deformed surface of water. When decreasing the water depth, we observe a drastic transition between weak turbulence at the weakest forcing and a solitonic regime at stronger forcing. We characterize the transition between both states by studying their Fourier Spectra. We also study the efficiency of energy transfer in the weak turbulence regime. We report a loss of efficiency of angular transfer as the dispersion of the wave is reduced until the system bifurcates into the solitonic regime. 
\end{abstract}

\maketitle

\section{Introduction}

Wave turbulence and solitonic turbulence are two major faces of the propagation of random and weakly nonlinear waves. The dynamics of incoherent dispersive waves has been studied in the framework of the statistical theory of Weak Turbulence since the pioneering work of Zakharov, Benney and Newell in the 60's (see \cite{Nazarenko,newell_wave_2011} for recent reviews). Based on the hypothesis of weak nonlinearity, the theory predicts an energy transfer through resonant interactions among waves. For forced turbulence it leads most often to a direct energy cascade from the forced large scales to small scales at which dissipation is acting. The Weak Turbulence Theory (WTT) provides analytical predictions of the statistical properties of turbulence such as the stationary wave spectrum. This theory has been applied to many physical systems such as plasmas \cite{sagdeev19791976}, optics \cite{picozzi2014optical}, thin elastic plates~\cite{During} or geophysical flows~\cite{Hasselmann,Galtier}. In particular, it was applied to wave propagating at the surface of deep water either in the gravity regime~\cite{Hasselmann} or the capillary regime~\cite{Filonenko}. In the last decades wave turbulence at the surface of a fluid has been the object of several detailed experimental investigations in either regimes or at the gravity-capillary crossover~\cite{Lukaschuk,Cobelli,Falconrev,aubourg_nonlocal_2015} that triggered a new dynamics in the field. 

For weakly non dispersive waves, weakly non linear waves may also develop into solitons i.e. localized structures that propagate at a constant velocity while keeping their shape unchanged~\cite{Dauxois}. 
They have been discovered by Russell~\cite{russell1845report} and modeled by the Korteweg-de Vries (KdV) equation~\cite{korteweg1895xli}. The interaction between solitons is elastic and appears as a phase shift between pre- and post- collision solitons. When a large number of solitons evolves in a medium, they develop a statistical state called soliton gas. If soliton gases have been experimentally observed for a long time in optics \cite{schwache1997properties,mitschke1999soliton}, their observation at the surface of fluids has been reported only recently~\cite{costa_soliton_2014,perrard2015capillary}. Soliton gases develop a quite different turbulent state than weak turbulence due to the integrability of the KdV equation~\cite{Zakharov_TIS} for which an infinite number of invariants exist in addition to energy.

Here we report the observation of a transition between both weak turbulence and a solitonic regime. We focus on the case of weakly nonlinear waves propagating at the surface of a water layer at scales typical of the gravity-capillary crossover (i.e. wavelengths at centimeter scale). In this configuration weak wave turbulence has been unambiguously reported for a deep enough layer and weak enough forcing using space and time resolved measurements~\cite{aubourg_nonlocal_2015,aubourg2016investigation,Cobelli}. Such measurements are required to accurately characterize the wave propagation and identify the nonlinear regimes. At stronger forcing another wave turbulence regime has been identified that is more strongly nonlinear and involves coherent structures~\cite{CobelliPrz,Berhanu}. Shallow water weak wave turbulence was also observed numerically by di Leoni {\it et al.}~\cite{di2014wave} for gravity waves. Furthermore at finite depth gravity-capillary solitons are known to propagate~\cite{falcon_observation_2003}. We report an investigation of the influence of the water depth and the intensity of the forcing on the statistical properties of wave turbulence. Changing the water depth alters the dispersivity of the waves. We observe and characterize the transition between a weak turbulence regime (low forcing and/or large depth) and a solitonic regime (larger forcing and/or small depth) that may evolve into a soliton gas.

\section{Experimental Set-up}

The experimental set-up is similar to that of Aubourg $\&$ Mordant~\cite{aubourg_nonlocal_2015}. It consists in a rectangular $57\times37$~cm$^2$ Plexiglas tank filled with water. The depth of water at rest is changed in the range $[0.6,6]$~cm, going from finite depth regimes to deep water regimes considering the centimetric wavelengths. Great care is taken to prevent surface contamination in order to avoid dissipation by conversion to Marangoni waves~\cite{Przadka}. The waves are generated by exciting continuously horizontally the container (along its longest axis) using an oscillating table controlled by a sinusoidal tension modulated around a central frequency equal to  2~Hz with a random modulation in an interval $\pm$ 0.5~Hz around the central frequency. The deformation of the free surface is reconstructed using the Fourier Transform Profilometry \cite{cobelli_global_2009,maurel_experimental_2009} (see reference \cite{aubourg2016investigation} for more details of the set-up). The water is mixed with titanium dioxide particles at a volume fraction of 1$\%$ to improve its optical diffusivity making possible to project on its very surface a grayscale sinusoidal pattern. These particles are  chemically neutral and do not alter the surface tension of water as stated by Przadka et al \cite{Przadka}. As the waves propagate the pattern seen by a high speed camera is deformed. The deformation of the pattern can be inverted to provide the elevation field of the waves. In this way we obtain a full space-time resolved measurement of the wave field. The images are recorded at 250 frames/s with a resolution of $1024\times1024$~pixels$^{2}$ covering a 20~cm$^{2}$ area at the center of the tank.

\begin{figure}[!htb]
\centering
(a)  \includegraphics[clip,width=8cm]{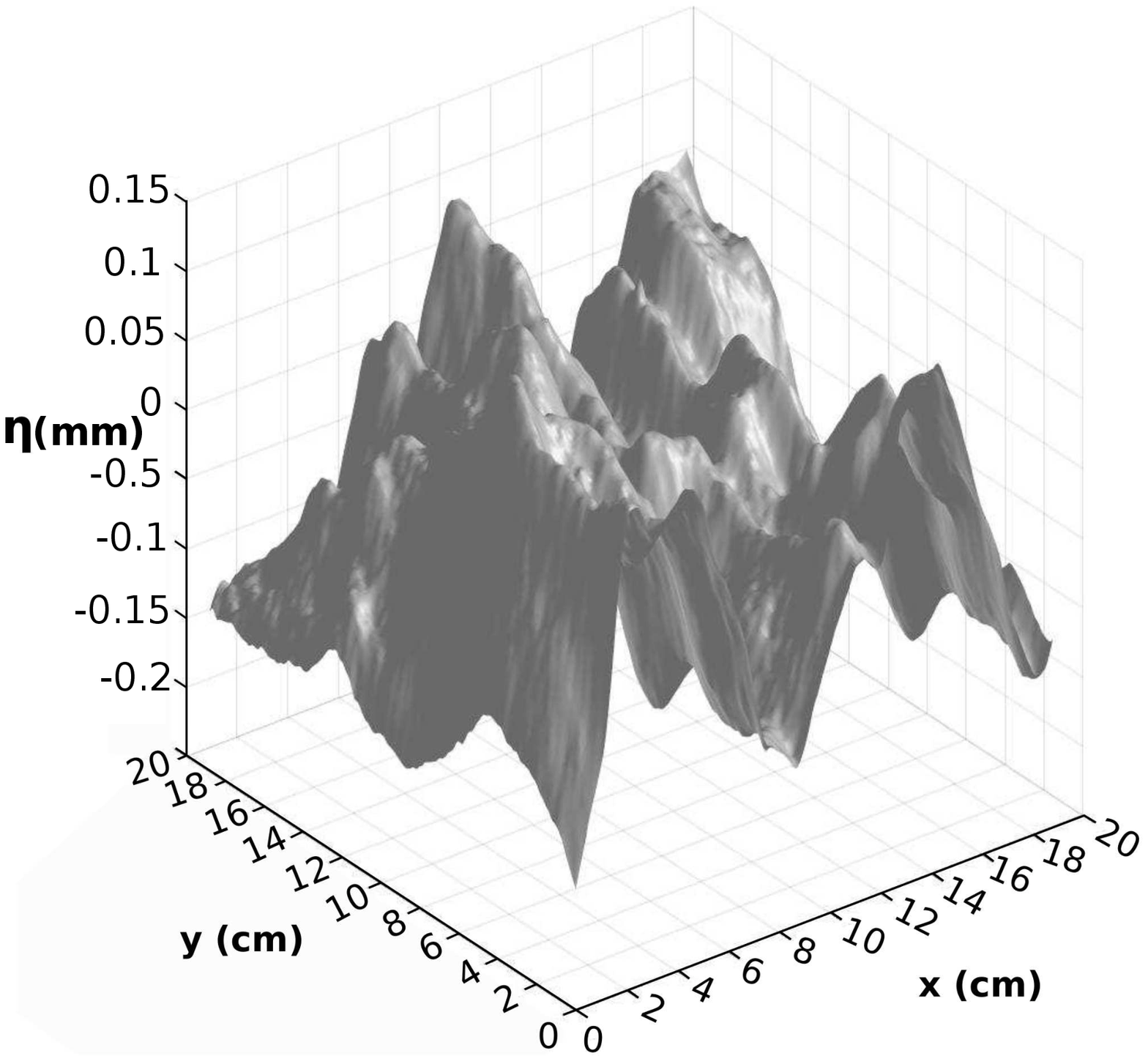} \hfill
(b)  \includegraphics[clip,width=8cm]{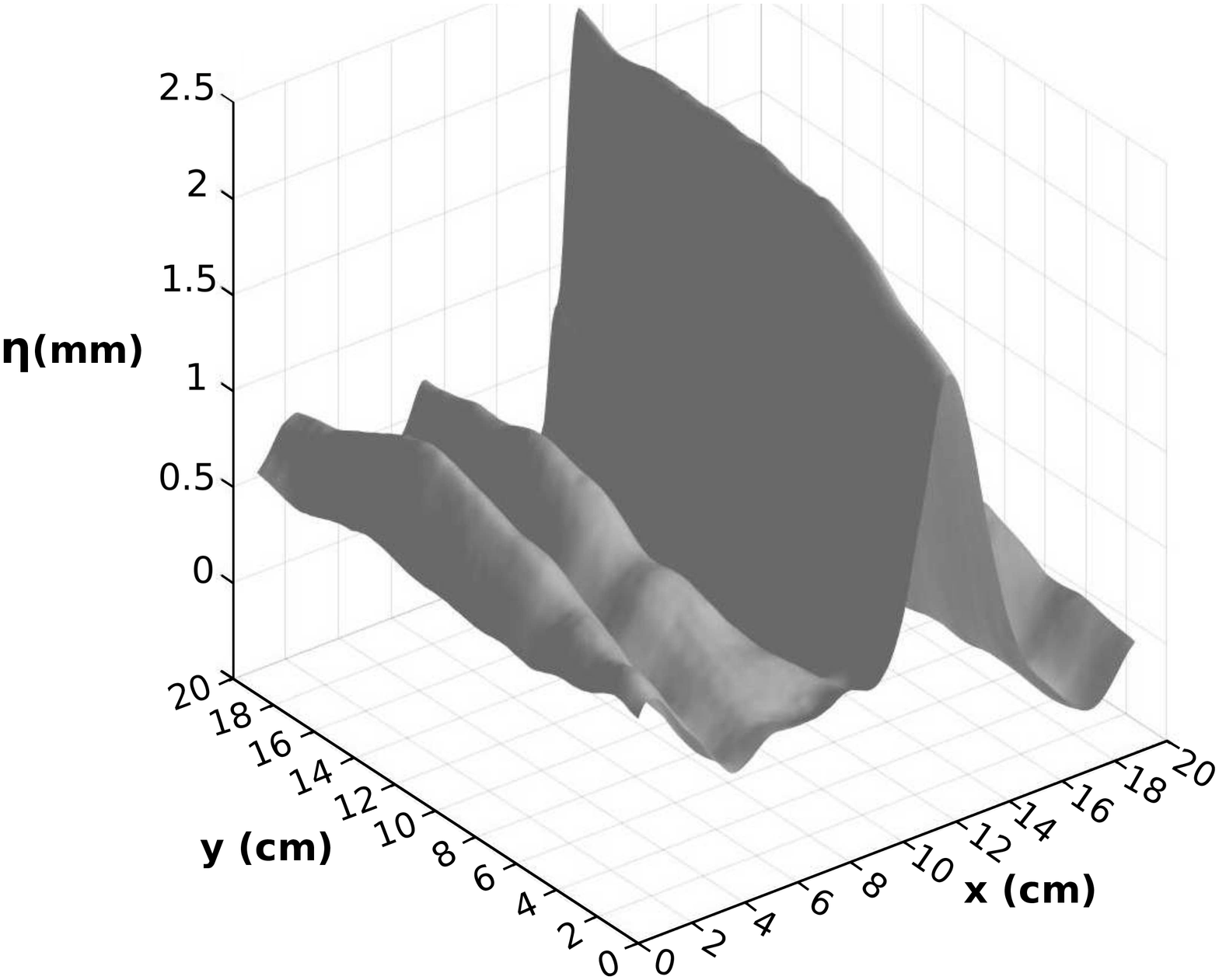}
\caption{Elevation field of the waves for a height of water at rest equal to 1~cm. The excitation central frequency is 2~Hz. (a)  amplitude of forcing equal to 0.9 mm. The {\it rms} slope is 1.6\% and this regime corresponds to wave turbulence. (b) amplitude of forcing equal to 2.5~mm. The {\it rms} slope is 4.1\% and a soliton can clearly be identified. In both cases the magnitude of the waves has been magnified strongly vertically.}
\label{surface}
\end{figure}

\begin{figure}[!htb]
\centering
\includegraphics[width=12cm]{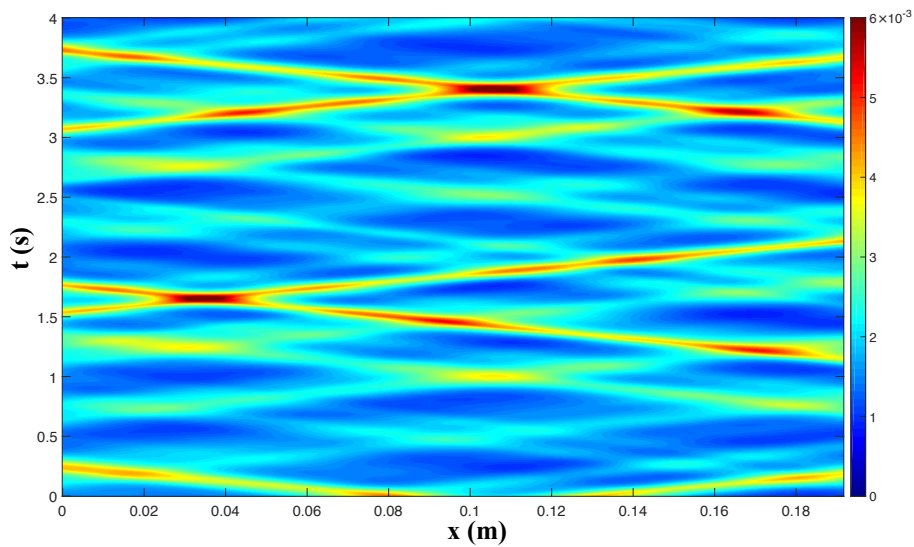} \hfill
\caption{Space-time plot of the solitonic regime of fig.~\ref{surface}(b). The excitation central frequency is 2~Hz and the amplitude of forcing equal to 2.5~mm. The colormap is the wave height in meters.}
\label{h_t_x}
\end{figure}

We perform experiments with various magnitudes of the forcing and several values of the water depth at rest.
Figure \ref{surface} shows the surface reconstruction in the case of a weak and a strong forcing amplitude.
At weak forcing, fig.~\ref{surface}(a) shows a random distribution of the waves. The deformation seems to involve waves coming from all directions. At stronger forcing (fig~\ref{surface}(b)) we observe a localized coherent structure. These soliton-like, coherent structures are only propagating along the direction of the oscillation of the table ($x$ axis). Note that there is about a factor ten between the amplitudes of the reconstructed waves in both cases. In this regime the system contains actually several solitons  which propagate in both directions along the $x$-axis and collide as represented in a time space representation in fig.~\ref{h_t_x}.

\section{Spatio-temporal spectral Analysis}

\begin{figure}[!htb]
(a)  \includegraphics[clip,width=8cm]{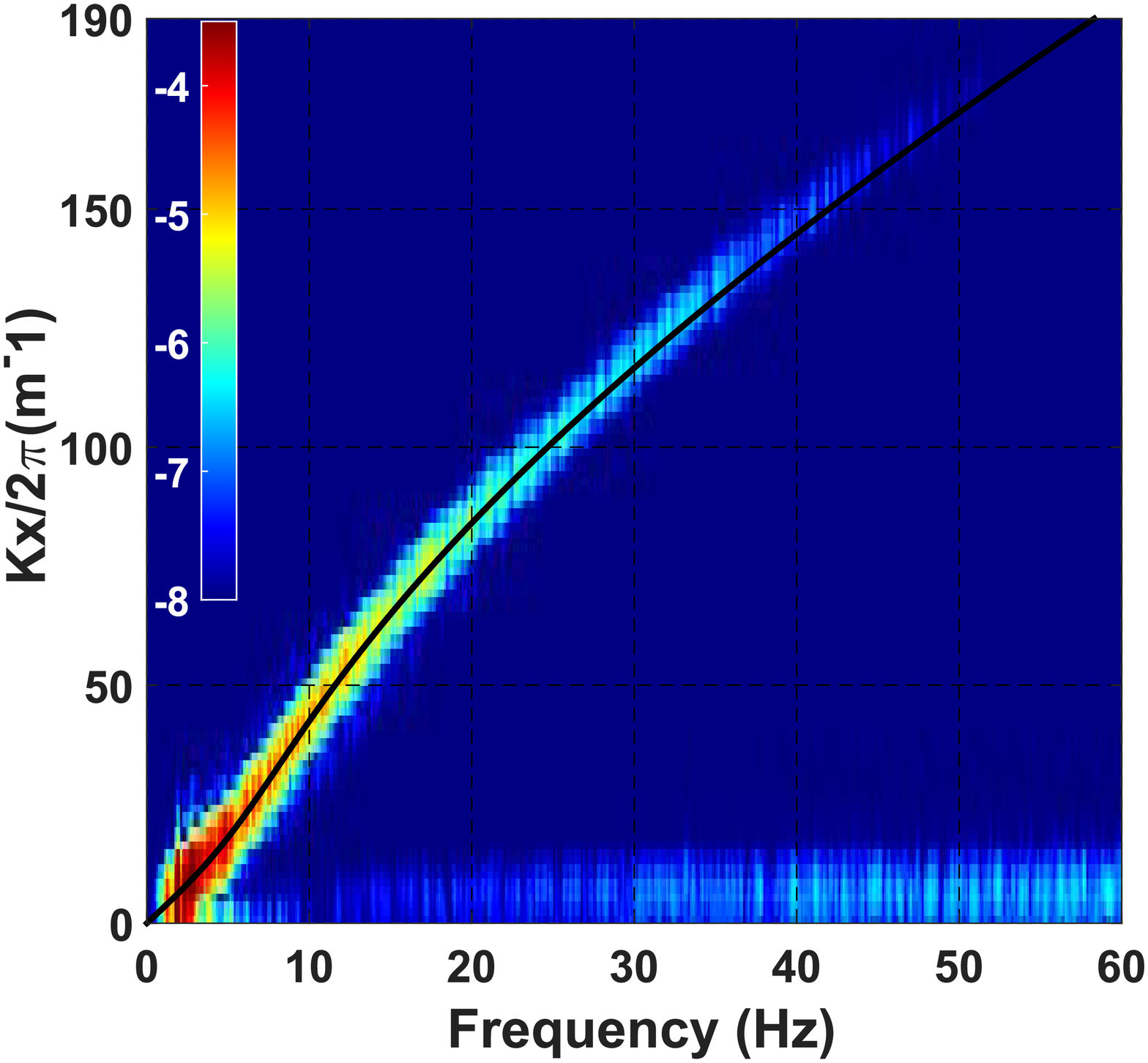} \hfill
(b)  \includegraphics[clip,width=8cm]{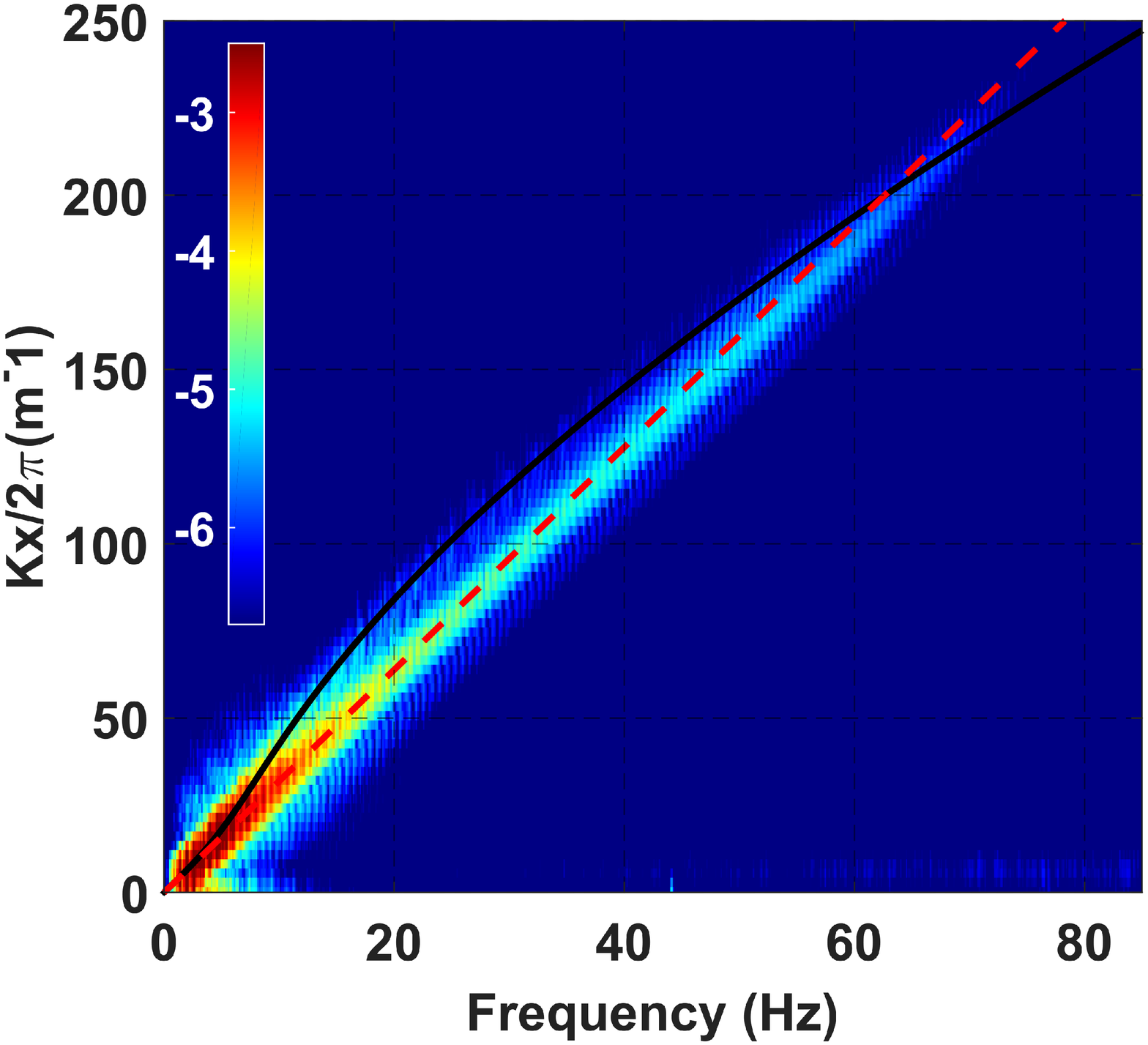}

\caption{Space-time Fourier spectrum $E^v(\mathbf k,\omega)=\langle |v(\mathbf k,\omega)|^2\rangle$ of the velocity field $v=\frac{\partial \eta}{\partial t}$ of the waves. The height of water at rest is 1~cm. The forcing frequency is centered around 2~Hz. (a) $E^v(k_x,k_y=0,\omega)$ for an amplitude of forcing equal to 0.9 mm. (b) same quantity for an amplitude of forcing equal to 2.5 mm. In both (a), (b), the solid black line is the theoretical finite depth linear dispersion relation for gravity-capillary waves in pure water with surface tension $\gamma=72$~mN/m. In (b) the dashed red line is a straight line of slope $1/c_{s}$ with $c_{s}=\sqrt{gh}$ the celerity of long waves. In both subfigures the color is the log of $E^v (\mathbf k,\omega)$. 
}
\label{spectra}
\end{figure}

\begin{figure}[!htb]
\includegraphics[clip,width=16cm]{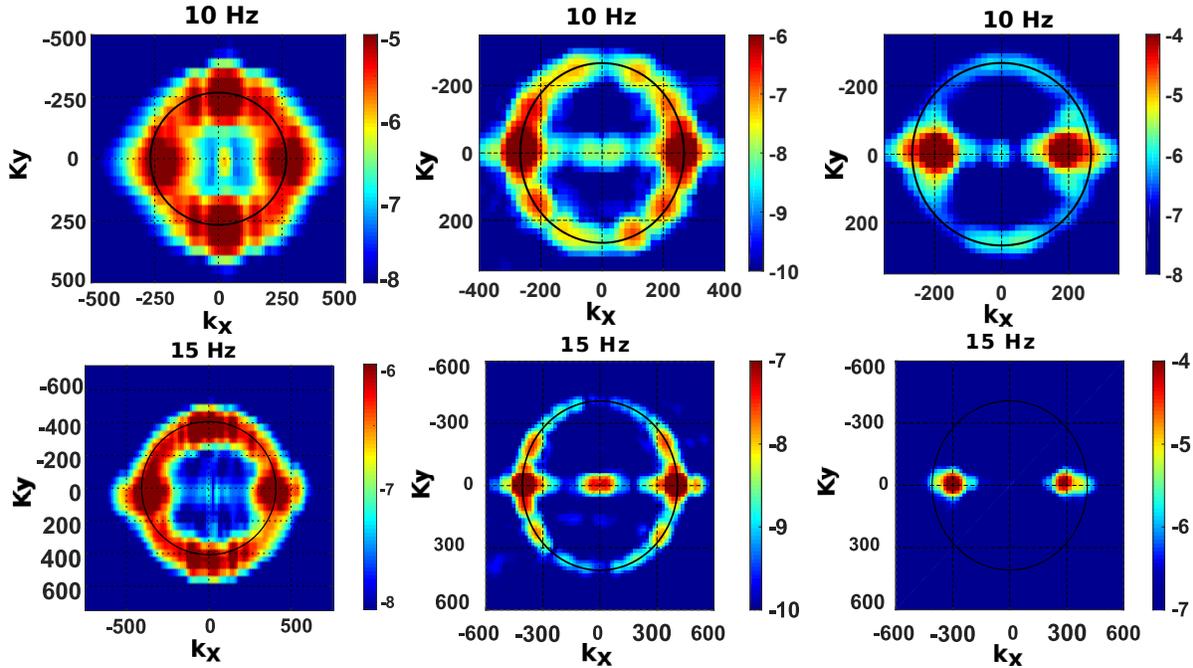}

\caption{Cut of the energy spectrum $E^v(k_{x},k_{y},\omega)$ at a given value of the frequency $F=\omega/2\pi$ and shown as a function of $(k_{x},k_{y})$. The solid black line corresponds to the finite depth water gravity-capillary dispersion relation~(\ref{gc_disp}). The top line  corresponds to $F=10$~Hz and the bottom line corresponds to $F=15$~Hz. The left column comes from a weakly non-linear regime in deep water $h=6$~cm where the system is quite isotropic, the central column comes from the WTT experiment in fig.~\ref{spectra}(a) and the right column corresponds to the soliton experiment of fig.~\ref{spectra}(b).}
\label{isotropy}
\end{figure}

In order to identify the spectral content of the wave field, we compute the full frequency-wavenumber Fourier spectrum. For a better visualization of the spectra, we actually compute the spectrum of $\frac{\partial \eta}{\partial t}$, which can be interpreted as the vertical velocity due to the weak magnitude of the nonlinearity ($\eta(x,y,t)$ is the instantaneous height of the free surface). The spectrum is noted $E^v(\mathbf k,\omega)$ and is shown in fig.~\ref{spectra} for the two experiments of fig.~\ref{surface}. At weak forcing (fig.~\ref{spectra}(a)), the energy is concentrated on the dispersion relation of gravity-capillary waves at finite depth:
\begin{equation}
\omega = \sqrt{\left(gk + \frac{\gamma}{\rho}k^3\right)\tanh(kh)}
\label{gc_disp}
\end{equation}
where $\gamma= 72$~mN/m is the surface tension of pure water, $\rho$ is the water density = $10^3$~kg/m$^{3}$, and $g=9.81$~m/s$^2$ is the acceleration of gravity. At large wavelength $kh\ll1$, $\omega\approx\sqrt{gh}k$ and the waves are weakly dispersive. The concentration of energy on this linear dispersion relation is the clear indication of the presence of weak turbulence as reported in~\cite{aubourg2016investigation,aubourg_nonlocal_2015}.

At stronger forcing (fig.~\ref{spectra}(b)), when localized structures are present, the energy is concentrated around a straight line of slope $1/c_s$ where $c_s=\sqrt{gh}$. $c_s$ is the velocity expected for solitons. A similar spectral signature has been also reported for solitons in nonlinear optics \cite{laurie2012one}. In our case the energy is concentrated mostly at large wavelength ($\lambda\gtrsim 2$~cm, corresponding to the size of the core of the soliton $\approx5$~cm) but it extends down to millimeter wavelengths due to the capillary precursors (reported for gravity-capillary solitons by Falcon {\it et al.}~\cite{falcon_observation_2003}) that are expected when the Bond number $Bo = \frac{\gamma}{\rho gh^2}$ lies between 0 and 1/3 (which is the case for all our experiments)~\cite{grimshaw1993solitary}. At low frequency the spectrum is tangent to the shallow water non dispersive part of the linear dispersion relation but departs clearly from it at smaller wavelengths. 

Figure~\ref{isotropy} displays cuts of the spectrum $E^v(\mathbf k,\omega)$ at constant frequency. For deep water ($h=6$~cm, left column), the spectrum is quite isotropic although the forcing is acting along the $x$ axis. This case is very similar to the one previously reported by Aubourg \& Mordant~\cite{aubourg2016investigation,aubourg_nonlocal_2015}. It shows a strong angular redistribution of energy. At lower depth and weak forcing ($h=1$~cm, weak turbulence regime, central column, shown in fig.~\ref{surface}(a) \& \ref{spectra}(a)), the spectra are more anisotropic with energy mostly in the $x$ direction with a much weaker angular spreading. At the same depth but stronger forcing (soliton regime, right column, shown in fig.~\ref{surface}(b) \& \ref{spectra}(b)), energy is only visible on the $x$ axis, along the direction of propagation of the solitons. Furthermore the peak of energy is clearly out of the linear dispersion as already observed in fig.~\ref{spectra}. A first consequence of reducing the dispersivity of the waves seems to make the angular transfer of energy less efficient until the transition to the soliton regime at which no angular transfer is observed.

\begin{figure}[!htb]
(a) \includegraphics[clip,width=8cm]{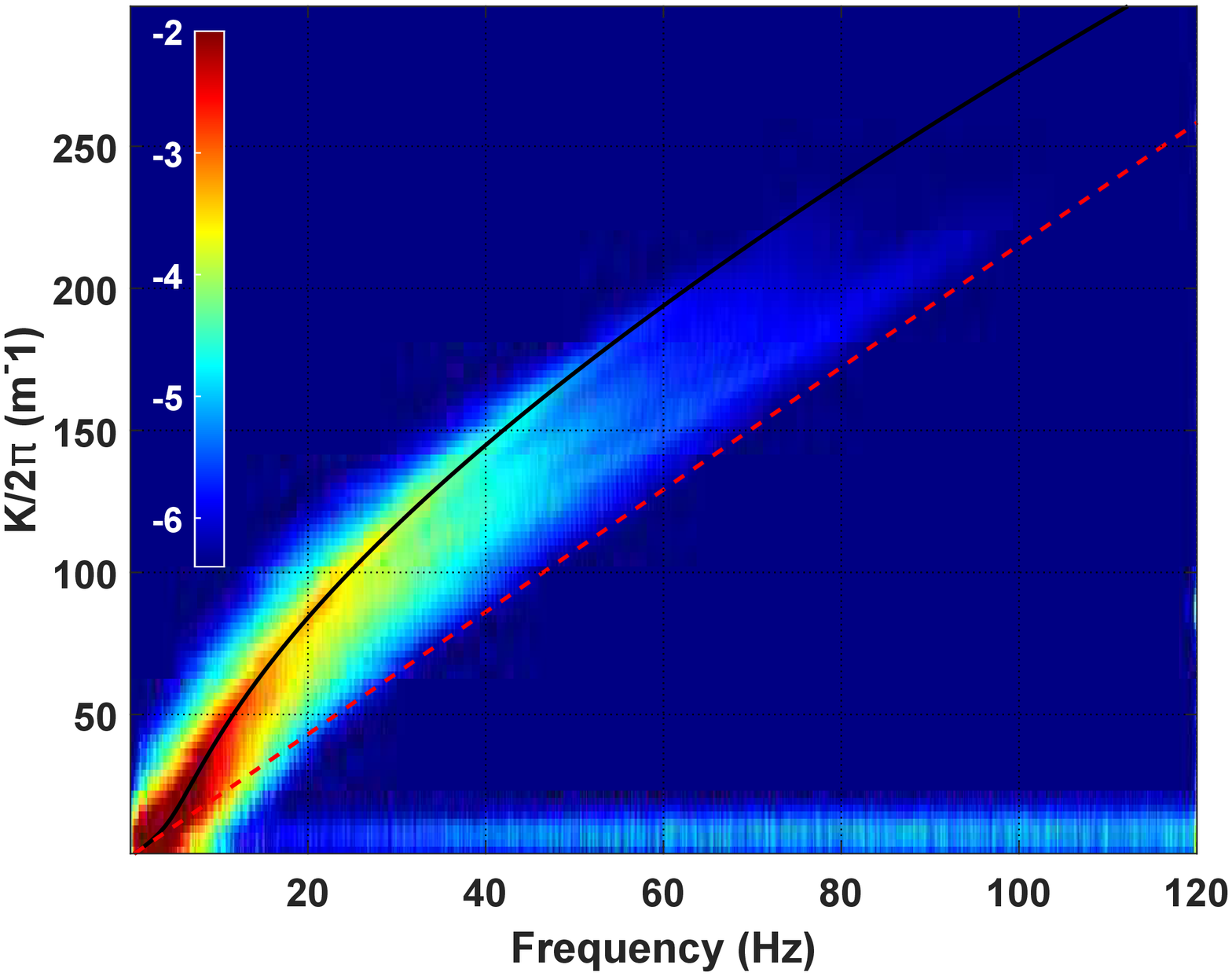}
(b) \includegraphics[clip,width=8cm]{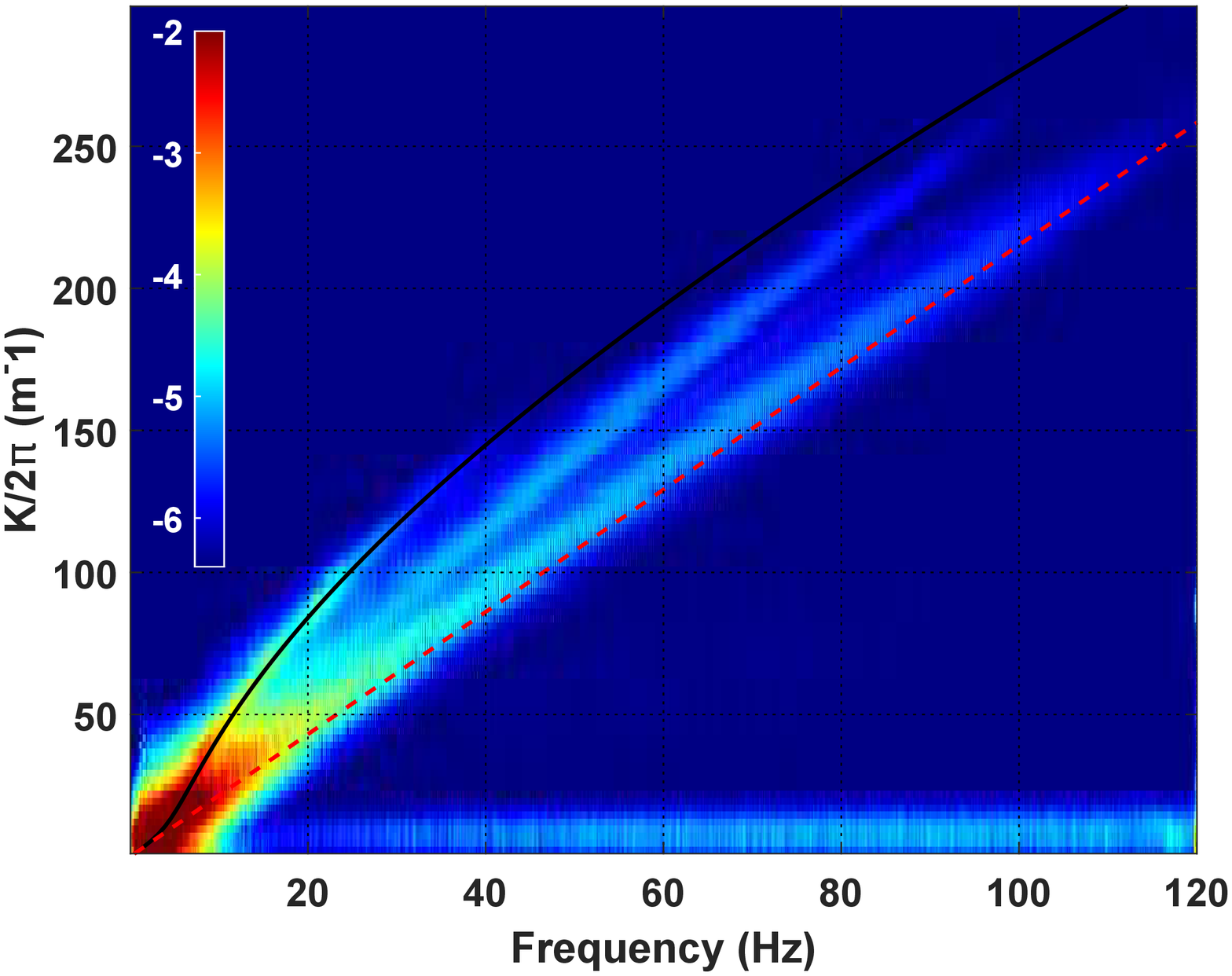}
\caption{Space-time Fourier spectrum $E^v(k,\omega)$ of the velocity field of the waves  for experiments where the height of water at rest $h =3$~cm with a forcing centered around 0.7~Hz in an intermediate regime between weak turbulence and solitonic regime. (a) amplitude of table oscillation equal to 6.6~mm. The {\it rms} slope is 6$\%$ (b) amplitude of table oscillation equal to 9.2~mm. The {\it rms} slope is 6.9$\%$. The solid black line corresponds to the gravity-capillary linear dispersion relation in finite depth for pure water. The red dashed line corresponds to the celerity of the soliton $c_{s}=\sqrt{gh}$.}
\label{transient}
\end{figure}

We also observe intermediate states between the WTT and the ST represented in fig.~\ref{transient} where some energy is observed in between the linear dispersion relation and the solitonic straight dispersion at the celerity $c_{s}=\sqrt{gh}$. This intermediate stage could be seen only for experiments with finite depths higher than 1 cm and a frequency of forcing lower than 2~Hz. In fig.~\ref{transient}(a) energy is observed on the linear dispersion relation and below but not yet on the soliton line. Extra energy lines can be seen notably in fig.~\ref{transient}(b) that correspond to non dispersive structures that propagate at a different velocity. This state corresponds to the beginning of the development of solitons that may not enough stable to resist the influence of the large scale stationary modes of the tank that possibly tend to make them loose their coherence very quickly. When the amplitude of the forcing is increased the solitonic state is then reached and energy is concentrated around the expected soliton dispersion relation.  

\begin{figure}[!htb]
\includegraphics[clip,width=9cm]{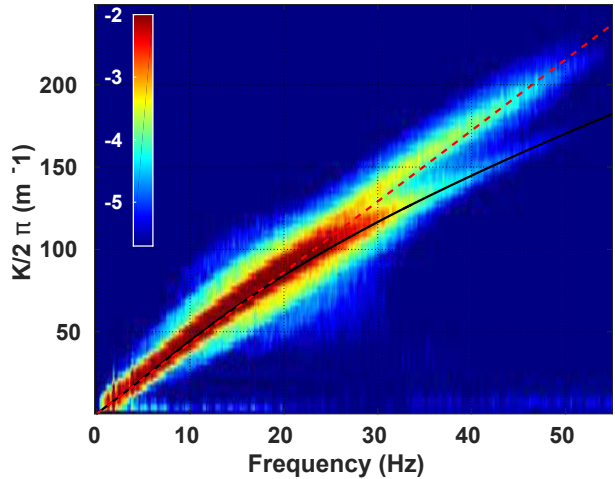}
\caption{ Space-time Fourier spectrum $E^v(k,\omega)$ of the velocity field of the waves for experiments where the height of water at rest is h = 6mm with a forcing centered around 2~Hz and an amplitude of forcing is 3~mm. The rms slope is measured to be 6.4$\%$. The solid black line corresponds to finite depth water gravity-capillary linear dispersion relation. The red dashed line corresponds to the celerity of the soliton $c_{s}=\sqrt{gh}$.}
\label{soliton_6mm}
\end{figure}
As shown in the previous paragraph, the spectral signature of solitons in $(k_x,\omega)$ space corresponds to a straight line that goes through the origin and is tangent to the shallow water gravity-capillary dispersion relation at the lowest frequencies. For a water depth equal to 1~cm (fig.~\ref{spectra}(b)) the line is located below the dispersion relation (for frequencies lower than 60~Hz). For a water depth equal to 6 mm, the energy line representing the soliton is above the dispersion relation as shown in fig.~\ref{soliton_6mm}.  This represents a transition between a soliton that has a higher celerity than linear waves to a soliton that is slower than linear waves. This is called the bifurcation between the subcritical branch and the supercritical branch as explained by Kuznetsov \& Dias in their review \cite{kuznetsov2011bifurcations}. Analytically the inflection point disappears when the Bond number reaches $1/3$, i.e. when $h=\sqrt{\dfrac{3\gamma}{\rho g}}=4.7$~mm for pure water). In fig.~\ref{soliton_6mm} the Bond number is slightly below $1/3$ so that the inflection point is barely visible (compared to the spread of energy around the dispersion relation). The dispersion relation is close to linear in a large interval up to about 12Hz. The soliton energy line is superimposed with the dispersion relation in this interval and then lies above it at higher frequencies. In the previous case of fig.~\ref{spectra}(b) the soliton energy is above the dispersion relation only above 60 Hz. In this configuration we are still in the subcritical case but we clearly see the evolution to the supercritical case.

\section{Frequency spectra}

For pure power law dispersion relations, the Weak Turbulence Theory predicts the expression of the energy spectrum $E(k)$ as : 

\begin{equation}
 E(k) = CP^{1/(N-1)}k^{-\alpha}
\label{spectrum_k}
\end{equation}

where $ k = \| \mathbf{k} \|$ is the wave number, $P$ the energy flux, $C$ a dimensional constant that can be calculated and $\alpha$ the spectral exponent. $N$ is the number of waves taking part in the non linear coupling~\cite{Nazarenko}. Using the dispersion relation, one can translate the expression of $k$ spectra into $\omega$ spectra that are often easier to measure. One of the outcomes of WTT is that energy is transmitted among resonant waves. At the lowest order it involves 3 waves that have then to satisfy the resonance conditions such as  $\mathbf{k_{1}}=\mathbf{k_{2}}+\mathbf{k_{3}}$  and  $\omega_{1}=\omega_{2}+\omega_{3}$. This is the case for pure capillary waves on infinite depth water~\cite{Filonenko}.
In the case of gravity waves the 3-wave resonance conditions do not admit solutions because of the negative curvature of the dispersion relation in infinite depth  $\omega = \sqrt{gk}$. One must take into account the next order that involves 4 waves. The predicted spectra of the surface deformation $\eta$ are thus expected to be 
\begin{equation}
E(k) \propto P^{1/3}gk^{-5/2}
\end{equation} 
for deep water pure gravity waves and 
\begin{equation}
E(k) \propto P^{1/2}k^{-7/4}
\end{equation}
for deep water pure capillary waves~\cite{Nazarenko}.
Using the dispersion relations for gravity waves and capillary waves in finite depth one can change the variable from $k$ to $\omega$ in the predicted spectra and obtain  
\begin{equation}
E(\omega) \propto P^{1/3}g\omega^{-4}
\end{equation} 
for gravity waves and 
\begin{equation}
E(\omega) \propto P^{1/2}\left(\frac{\gamma}{\rho}\right)^{1/6}\omega^{-17/6}
\end{equation}
for capillary waves. Laboratory experiments fail to reproduce this predictions for the gravity waves. In large or small waves tanks, the spectral exponent of the gravity waves is seen to vary strongly with the forcing intensity and to be close to the WTT predictions only at the highest forcing magnitude, in opposition to the weak nonlinearity hypothesis \cite{nazarenko2010statistics,falcon2007observation}. For capillary waves, experiments seem to follow better the predictions of the WTT \cite{wright1997imaging,deike2012decay}.


\begin{figure}[!htb]
\includegraphics[clip,width=11cm]{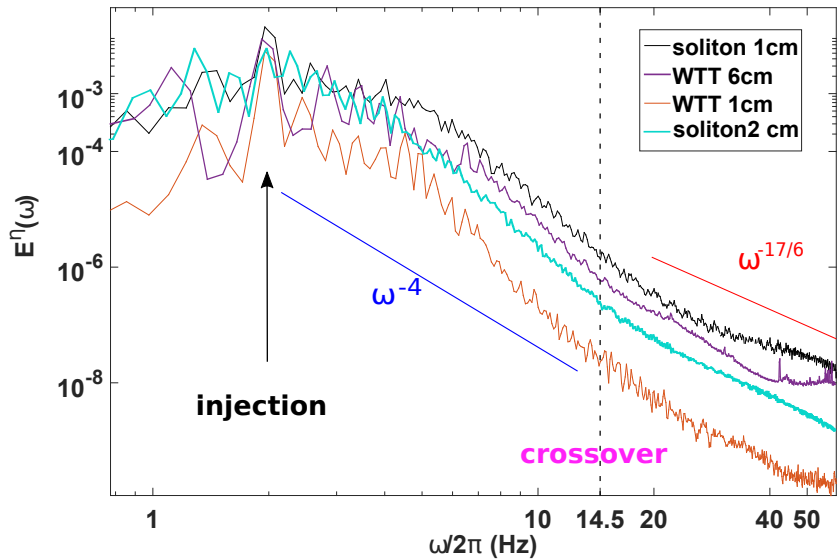}
\caption{Time Fourier spectrum $E^v(\omega)$ of the surface deformation for experiments where the height of water at rest is 1~cm with a forcing centered around 2~Hz and an amplitude of forcing equal to 2.5~mm. The vertical dashed line corresponds to the frequency of the crossover between the gravity and the capillary regime for pure water. The injection scale corresponds to the frequency of the forcing.}
\label{fspec}
\end{figure}

Figure~\ref{fspec} shows the frequency spectra for various experimental configurations. The spectra corresponding to a weak turbulence regime (at 1~cm and 6~cm depth) are very similar to that of Aubourg \& Mordant~\cite{aubourg2016investigation}. At frequencies below the gravity-capillary crossover (at 14.5~Hz) and thus corresponding to gravity waves the spectra are steeper than the WTT prediction, the steepest being observed at low depth. In the capillary range, the spectra are also steeper than the WTT prediction consistently with Aubourg \& Mordant~\cite{aubourg2016investigation} but at odds with observation by Falcon {\it et al.}~\cite{Falconrev} who observed a $\omega^{-3}$ scaling close to the WTT prediction. In the solitonic regime, the spectrum is not changed much in the gravity range. In the capillary range the spectra are changed and are closer to the WTT prediction and even less steep at 1~cm depth. This may suggest that the $\omega^{-3}$ scaling observed by Falcon {\it et al.} may correspond to the strong regime reported by Cobelli et al.~\cite{CobelliPrz}. In their figure 1 (top, strong regime) the surface velocity shows coherent structures that may be some sort of gravity-capillary solitons generated by the paddles even if their water depth is not that small (5~cm). The $\omega^{-3}$ scaling may be due to the solitons as it is the case in our experiment at 2~cm water depth rather than being due to a fulfillment of the WTT prediction.

\section{Phase Diagram}

The solitons that have been observed in our work are expected to follow the Korteweg-de Vries equation:
\begin{equation}
 \eta_{t} + \frac{3}{2}\frac{c_{s}}{h} \eta\eta_{\xi} + \frac{1}{6}c_{s}h^{2}(\frac{1}{3} - Bo)\eta_{\xi\xi\xi} = 0 
\label{kdv}
\end{equation}
with $c = c_{s}( 1 + \frac{\eta_{0}}{2h})$ and $\xi = x -ct$. Note that $ \eta_{0} < 0$ when $Bo > \frac{1}{3}$ , $\eta_{0} > 0$ when $0 \leq Bo < \frac{1}{3}$.
Equation \eqref{kdv} can be derived only if nonlinear effects are small and have the same order of magnitude as dispersive ones.
Dispersion is quantified by $\mu = (\frac{h}{L})^{2}$ and the non-linearity is quantified using the typical {\it rms} steepness of the waves
\begin{equation}
\epsilon \equiv \sigma = \left\langle \sqrt{\frac{1}{S}\int_S \parallel \nabla \eta(x,y,t) \parallel ^{2}\,\mathrm dx dy} \right\rangle \, .
\end{equation}

\begin{figure}[!htb]
\center
\includegraphics[width= 10cm]{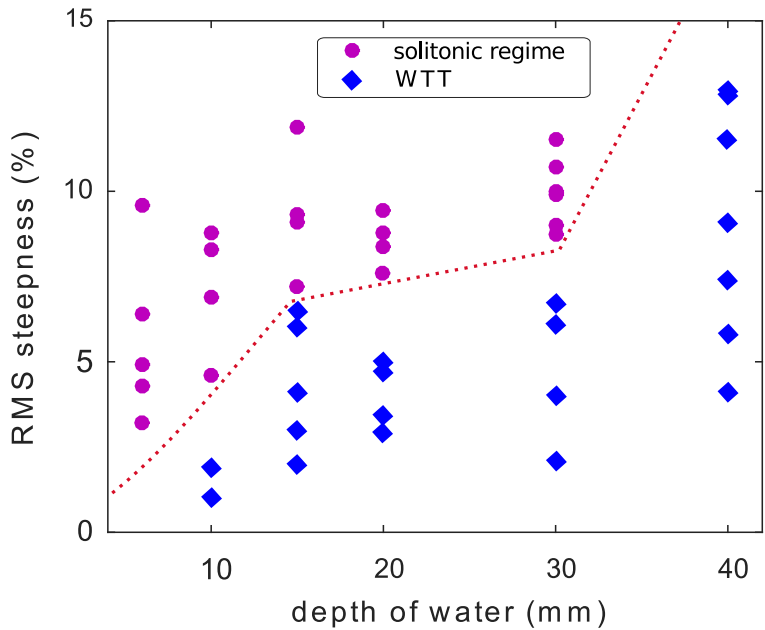}
\caption {Solitonic transition phase diagram represented as a function of the {\it rms} steepness of the waves and the depth of water at rest, for experiments where the central frequency of excitation is 2~Hz. The dashed red line delimits the area where localized structures are predominant from the one where random waves are predominant}
\label{A_h_2Hz}
\end{figure}
Figure \ref{surface} and \ref{spectra} pointed out major changes in the spatial and spectral signatures of the studied states where only the amplitude of the forcing was tuned. Figure~\ref{A_h_2Hz} displays a phase diagram of the weak turbulence regime and the solitonic one as a function of the water depth and the measured steepness of the waves. The central frequency is kept constant equal to 2~Hz for all the experiments. At a given water depth, the central frequency imposes indirectly the size of the soliton although predicting the actual size involves computing the inverse scattering transform of the excitation~\cite{trillo2016experimental}. This computation is beyond the scope of this article but we observe that the size of the soliton changes only weakly with the experimental parameters. Thus the water depth is a measure of the dispersivity of the waves while the steepness is a measure of the nonlinearity. Each case is categorized  as either solitonic or weakly turbulent by the observation of the Fourier spectrum. A border is clearly observed between both regimes. The border is increasing with the water depth: for deeper water it requires to force stronger waves to observe solitons. Equality between nonlinearity and dispersion is expected at the border, since it is the required condition to the development of solitary waves. To verify this condition we display in the table \ref{tableau_soliton} the minimal values of non-linearity and dispersion for which solitary waves were observed for different values of the water depth. Globally we observe that the balance between nonlinearity and dispersion is fullfilled at each depth of water. Dispersion is defined as $\mu = (\frac{h}{L})^{2}$ where L is the length of the core of the soliton. As $L$ is weakly changing it means that when the depth of water is increased the dispersion is stronger. The condition of balance of dispersion and nonlinearity imposes that the nonlinearity should be stronger as well in order to observe the solitons. In the deepest case ($h=4$~cm) it was not possible to observe solitons as the nonlinearity was becoming very high and the waves where close to overturning.
\begin{table}[!htb]
\center
\begin{tabular}{|l|c|c|c|c|c|c|r|}
  \hline
    h (cm) & 0.6 & 1 & 1.5 & 2 & 3 & 4 \\
  \hline
  $\epsilon$ (\%) & 3.2  & 4.6 & 7.3 & 7.6 & 8.8 & $\nexists$ \\
  $\mu$ (\%) & 3.8 & 4 & 7.2 & 7.3 & 9 & $\nexists$  \\
  \hline
\end{tabular}
\caption{Values of the nonlinearity $\epsilon$ and the dispersion $\mu$ for different depths of water with a frequency of excitation $\in$ $\left[1.5 \, \, 2.5\right]$~Hz. $\epsilon$ is estimated by directly measuring the surface averaged {\it rms} steepness of the waves. $\mu = (\frac{h}{L})^{2}$ is computed by measuring the depth of water at rest $h$ and $L$ the typical size of the core of the soliton}
\label{tableau_soliton}
\end{table}

\begin{figure}[!htb]
\center
\includegraphics[width= 10cm]{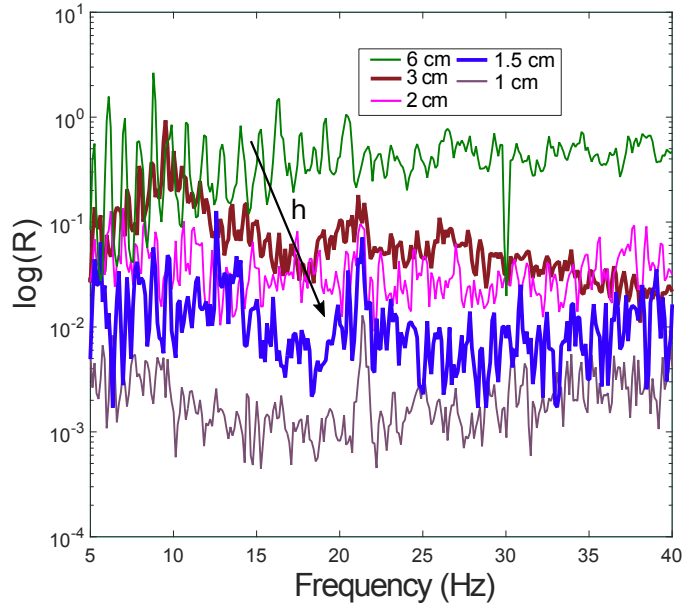}
\caption {Evolution of the ratio $R(\omega) = \frac{E^{v}(k_x=0,k_y(\omega))}{E^{\upsilon}(k_x(\omega),k_y=0)}$ of the energy perpendicular to the direction of the forcing oscillation over the energy in the direction of the forcing and for different depths $h$ of water. The energy level is taken on the linear dispersion relation $k(\omega)$. The central frequency of excitation is equal to 2 Hz. The {\it rms} steepness of the surface for values of the water depth $h = [ 6,\, 3,\, 2,\, 1.5,\, 1]$~cm are respectively $[ 4.4,\, 6.5,\, 5,\, 6.4,\, 4.1]\%$. In all cases the wave steepnesses are thus close to 5\%.}
\label{Et_El}
\end{figure}

\section{Angular Energy Transmission}

Although the transition between the two regimes seems very abrupt, the observation of the wave spectrum showed a progressive change in the angular transfer of energy  in the weak turbulence regime when decreasing the water depth. For truly non dispersive waves (acoustic waves), the resonant manifold involves only waves propagating in the same direction. Thus Newell \& Rumpf asked the question of the evolution of an initially anisotropic distribution of waves~\cite{newell_wave_2011}. Would higher order terms lead to angular transfer of energy or would the energy be condensed on rays that would evolve into shocks ? Here we decrease progressively the dispersion of the waves and we clearly see that the efficiency or angular transfer is altered. In order to be more quantitative we define $R$ as the ratio of the energy of the waves transverse to the forcing ($k_x=0$) divided by the energy of longitudinal waves propagating in the direction of the forcing ($k_y=0$):
\begin{equation}
R(\omega) = \frac{E^{v}(k_x=0,k_y(\omega))}{E^{\upsilon}(k_x(\omega),k_y=0)} 
\label{R}
\end{equation}
where $k(\omega)$ is the linear dispersion relation. Aubourg \& Mordant~\cite{aubourg_nonlocal_2015} reported for deep water that, although the energy injection is strongly anisotropic by forcing mostly waves in the $x$ direction, the nonlinear interactions between the waves redistribute isotropically this energy in a very efficient way. We show the evolution of $R(\omega)$ in fig.~\ref{Et_El}. We chose datasets for which the wave steepness (measure of the nonlinearity) is almost the same and close to 5\% in order to show only the effect of changing the dispersion of the waves. Consistently with the observation of Aubourg \& Mordant, $R$ is close to one in the deepest case. $R$ decays strongly when the water depth evolves from 6 down to 1~cm. This decay is roughly the same at all frequencies. When the water depth is $1.5$~cm, the ratio $R$ decayed by a factor 100. At 1~cm water depth the system transited into the soliton regime. In that regime the ratio is down to $10^{-3}$ showing that almost no directional transfer of energy is occurring. Answering Newell \& Rumpf, for finite depth gravity-capillary waves no angular transfer is induced by higher order closure but the system develops solitons rather than shocks as a weak dispersion remains present in our case.

\section{Conclusions}

In summary, we investigated the impact of nonlinearity and wave dispersion on wave turbulence. For strong dispersion of the waves (large water depth) and weak nonlinearity we observe a state of weak turbulence. When the dispersion of the waves is weak, the strength of nonlinearity can match that of dispersion. In these conditions, we observe the generation of solitons and thus a change of statistical regime from weak turbulence to soliton gas. The fate of energy is thus expected to be quite different. For weak turbulence, through resonant interactions, energy cascades to very small scales at which it is dissipated. In absence of dissipation, solitons are propagating without changing their shape and their interaction is elastic. An ensemble of solitons evolves into a state of integrable turbulence~\cite{Zakharov_TIS, Randoux} which properties are distinct from that of weak turbulence. Although in fiber optics the soliton gas is restricted to 1D propagation, for gravity-capillary waves solitons can have a 2D propagation and this maybe the case of the state observed by Cobelli {\it et al.}~\cite{Cobelli} at their strongest forcing. Note that Zakharov {\it et al.} predicted in \cite{zakharov_one-dimensional_2004} that for unidirectional propagation of gravity waves in deep water, weak turbulence should be unstable and coexist with soliton-like structures. In our case no coexistence of solitons and weak turbulence has been observed except very near the transition. We observe two very distinct regimes.

\begin{acknowledgements}
This project has received funding from the European Research Council (ERC) under the European Union's Horizon 2020 research and innovation programme (grant agreement No 647018-WATU). We thank V. Govart for his technical assistance.
\end{acknowledgements}

\bibliography{mabiblio}
\end{document}